# Lessons from Mayan Astronomy

A vibrant scientific culture should cultivate multiple approaches to analyzing existing data and to collecting new data, says **Avi Loeb.**

Recently, my family and I visited the Mayan city of *Chichen Itza* in Mexico, one of the New7Wonders of the world. As our tour guide marveled on the spectacular astronomical measurements of the Mayan culture, I began to wonder why our modern scientific understanding of Astronomy did not originate in South America. After reading some historical background, I realized that unfortunately the Mayans had used their exquisite astronomical data within a mythological culture of astrology that rested upon false but mathematically sophisticated theories about the Universe. They collected unprecedented amounts of precise astronomical data on the Sun, the moon, the planets and the stars, but failed to come up with the breakthrough ideas of Nicolaus Copernicus, Galileo Galilei, Johannes Kepler and Isaac Newton. The sober realization that scientific advances can be trapped by cultural and societal forces inspired me to wonder: *have we learned the necessary lessons to prevent our current scientific culture from resembling Mayan Astronomy?*

Let me sharpen the question further: *is data collection by itself a guarantee for good science?* This issue has important implications regarding the effectiveness of allocating resources by private and federal funding agencies for the advancement of science. Based on the historical success of empirically-based deduction in modern science, one would naively expect that a culture dedicated to the collection of the highest quality data would inevitably arrive at the proper scientific interpretation of this data and unravel the appropriate theory for making predictions that are testable by future data. With that as our model, all we would need to do in order to cultivate a productive scientific endeavor would be to fund state-of-the-art experimental or observational programs and scientific advances will simply follow. This popular strategy in funding science currently guides the allocation of most of the Astronomy Division funds at the US National Science Foundation (NSF) to major facilities and large scale surveys. The focus is clearly on large team efforts to collect better data within the mainstream paradigms of Astronomy, under the assumption that good science will follow.

The Mayan example leads to the realization, however, that a successful scientific culture requires a separate guiding principle, namely the desire to compare multiple interpretations of existing data and multiple motivations for collecting new data. This additional principle fosters scientific debate among competing theories, and encourages free thought outside the mainstream. Without this added principle, a culture could be scientifically inefficient at interpreting existing data and misguided in its motivation for collecting new data.

The Mayan culture collected exquisite astronomical data for over a millennium with the false motivation that such data would help predict its societal future. This notion of astrology prevented the advanced Mayan civilization from developing a correct scientific interpretation of the data and led to primitive rituals such as the sacrifice of humans and acts of war in relation to the motions of the Sun and the planets, particulary Venus, on the sky.

Fast forward another millennium. Cosmologists are currently collecting exquisite data within the paradigm of a single theoretical model in which the Universe contains a cosmological constant (Lambda) and cold dark matter (CDM), the so-called LCDM model. Heavily funded surveys are currently underway to pin down parameters of LCDM at percent level precision, and popular extensions of LCDM involve unsubstantiated but mathematically sophisticated notions of the multiverse, anthropic reasoning and string theory. Very often the vast data collected in large cosmological surveys is reduced to a few numbers, while possible surprises in the rest of the data are tossed away since they are not part of the programmatic agenda of the related team projects. I noticed this bias from close distance recently while serving on the PhD thesis committee of a student who was supposed to test whether a particular data set from a large cosmological survey is in line with LCDM; when a discrepancy was found, the goal of the thesis shifted to explaining why the data set is biased and incomplete. How can LCDM be ruled out in such a scientific culture? Observers should strive to present their results in a theory-neutral way rather than aim to reinforce the mainstream view.

The way each culture views the universe is often guided by auxiliary principles such as mathematical beauty or philosophical pre-notions about the structure of reality. If the notions are rooted deeply in our culture, we tend to interpret any data as supporting these notions by adding parameters or adopting mathematical gymnastics that accommodate the data within our conceptual infrastructure. The geocentric notion that the Sun moves around the Earth gave birth to the beautiful mathematical theory of epicycles advocated by Ptolemy. Similarly, the Mayans adopted the notion that the outcome of wars depended on the position of Venus in the sky, and so they correlated their societal history with accurate monitoring of the sky for the practical benefit of being able to forecast the future. Needless to say that such a world model tends to reassure itself by producing self-fulfilling prophecies, and accommodating contradictory empirical evidence by invoking complicated circumstances after the fact.

Today, mainstream physicists use the multiverse to explain the cosmological parameters of LCDM in the context of the advanced mathematics of the string theory landscape. The popular notion of the multiverse postulates the existence of numerous other regions of spacetime (to which we have no access as observers) in which the cosmological parameters obtain different values than in LCDM. The popular anthropic argument is that our own observable region possesses the LCDM parameters, and in particular the observed value of the cosmological constant, because other more likely values would not allow life to develop near a star like the Sun in a galaxy like the Milky Way[1-3]. One of the problems with this argument that was overlooked so far, is that life would be a thousand times more likely to exist around low mass stars (with a tenth of the mass of the Sun) in the distant future of our own region, ten trillion years from now[4]. The anthropic argument, which is founded on mathematical concepts from string theory but has no empirical support as of yet, suppresses much needed efforts to understand the true physical meaning of the cosmological constant in the context of an alternative theory that unifies quantum mechanics and gravity. The possibility that we have not converged yet on the final unified theory is indicated by additional conceptual problems that we encounter in other contexts, such as the information paradox of black holes[5] and the fine-tuning required for surviving models of inflation in light of the latest Planck data[6]. The multiverse notion is reminiscent of the mythological notions of the Mayans, which were not founded on empirical evidence but were formulated abstractly in the framework of the most advanced forms of mathematics and geometry of their time.

Given the strong sociological trends in the current funding climate of team efforts, how could we reduce the risk of replicating the indoctrinated Mayan astronomy? The answer is simple: by funding multiple approaches to analyzing data and multiple motivations to collecting new data. After all, the standard model of cosmology is merely a precise account of our ignorance: we do not understand the nature of inflation, the nature of dark matter or dark energy. Our model has difficulties accounting for what we see in galaxies (attributed often to complicated "baryonic physics"), while at the same time not being able to see directly what we can easily calculate (dark matter and dark energy). The only way to figure out if we are on the wrong path is to encourage competing interpretations of the known data.

Funding agencies should promote the analysis of data for serendipitous (non-programmatic) purposes. When science funding is tight, a special effort should be made to advance not only the mainstream dogma but also its alternatives. To avoid stagnation and nurture a vibrant scientific culture, a research frontier should always maintain at least two ways of interpreting data so that new experiments will aim to select the correct one. A healthy dialogue between different points of view should be fostered through conferences that discuss conceptual issues and not just experimental results and phenomenology, as often is the case currently. These are all simple, off-the-shelf remedies to avoid the scientific misfortune of the otherwise admirable Mayan civilization.

________________________________________________________________

**Avi Loeb** *is the* Frank B. Baird Jr. Professor of Science *and chair of the Astronomy department at Harvard University in Cambridge, Massachusetts, USA .*
*e-mail: aloeb@cfa.harvard.edu*

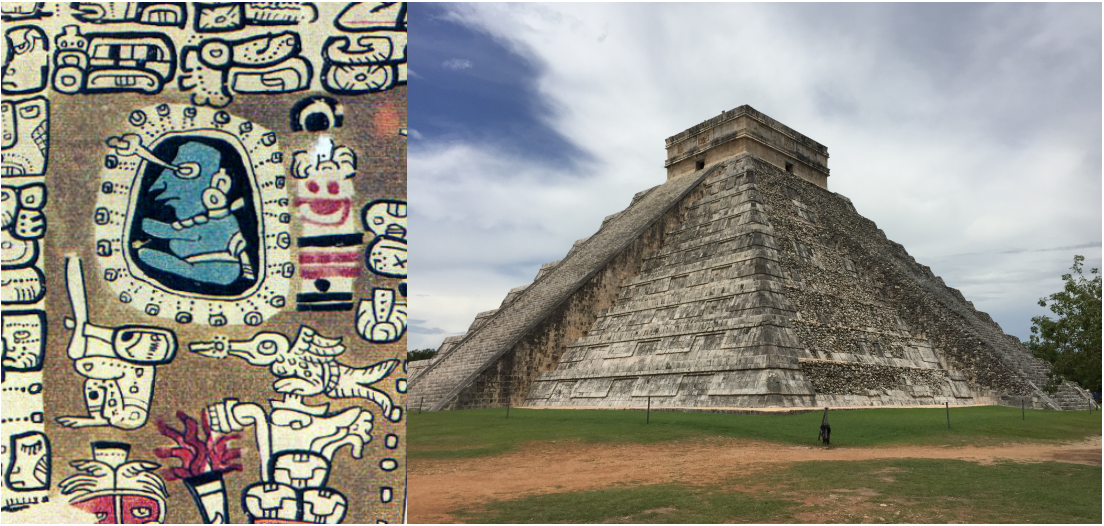

**Left:** A Maya astronomer, as depicted in the Maya book known as the "Madrid codex".
**Right:** El Castillo, also known as the temple of Kukulcan, at the Mayan city of Chichen Itza in Mexico (Photo credit: A. Loeb 2016). The Mayan culture involved a wonderful mix of state-of-the-art astronomical measurements and mathematics (in the spirit of our current mix of LCDM along with mathematical notions of the multiverse, anthropic reasoning and string theory). The 91 steps on each of the pyramid's sides together with its top complete a total of 365 steps, equal to the number of days in a year.